\documentclass[aps,twocolumn]{revtex4}        
\usepackage{amsmath,amssymb}

\begin{document}
\title{Friedmann equations from emergence of cosmic space}
\author{Ahmad Sheykhi\footnote{
asheykhi@shirazu.ac.ir}}
\address{Physics Department and Biruni Observatory, College of
Sciences, Shiraz University, Shiraz 71454, Iran\\
Center for Excellence in Astronomy and Astrophysics (CEAA-RIAAM)
Maragha, P. O. Box 55134-441, Iran}

\begin{abstract}
Padmanabhan [arXiv:1206.4916] argues that the cosmic acceleration
can be understood from the perspective that spacetime dynamics is
an emergence phenomena. By calculating the difference between the
surface degrees of freedom and the bulk degrees of freedom in a
region of space, he also arrived at Friedmann equation in flat
universe. In this paper, by modification his proposal, we are able
to derive the Friedmann equation of the Friedmann-Robertson-Walker
(FRW) Universe with any spatial curvature. We also extend  the
study to higher dimensional spacetime and derive successfully the
Friedmann equations not only in Einstein gravity, but also in
Gauss-Bonnet and more general Lovelock gravity with any spacial
curvature. This is the first derivation of Friedmann equations in
these gravity theories in a nonflat FRW Universe by using the
novel idea proposed by Padmanabhan. Our study indicates that the
approach presented here is enough powerful and further supports
the viability of the
Padmanabhan's perspective of emergence gravity.\\

PACS number:04.20.Cv, 04.50.-h, 04.70.Dy
\end{abstract}

\maketitle

\section{Introduction\label{Intr}}
Physicists have been speculating on the nature and origin of
gravity for a long time. Newton believed that gravity is just a
force like other forces of the nature and does not affect on the
space. This was a general belief until Einstein presented his
theory of general relativity in $1915$. According to Einstein's
theory, gravity is just the spacetime curvature. In this new
picture, the matter field tells space (geometry) how to curve, and
the geometry tells matter how to move. Also, according to the
equivalence principle of general relativity, gravity is just the
dynamics of spacetime. This implies that gravity is an emergent
phenomenon.

In $1970's$ thermodynamics of black holes were studied. According
to laws of black holes mechanics, a black hole can be regarded as
a thermodynamical system which has temperature proportional to its
surface gravity and an entropy proportional to its horizon area.
This indicates that geometrical quantities such as horizon area
and surface gravity are closely related to the thermodynamic
quantities like temperature and entropy. Are there a direct
connection between gravitational field equations describing the
geometry of spacetime and the first law of thermodynamics?
Jacobson \cite{Jac} was indeed the first who answered this
question by disclosing that the Einstein field equations can be
derived by applying the Clausius relation $\delta Q=T\delta S$ on
the horizon of spacetime, here $\delta S$ is the change in the
entropy and $\delta Q$ and $T$ are, respectively, the energy flux
across the horizon and the Unruh temperature seen by an
accelerating observer just inside the horizon.

The next great step toward understanding the nature of gravity put
forwarded by Verlinde \cite{Ver} in $2010$ who claimed that
gravity is not a fundamental interaction but should be interpreted
as an entropic force caused by changes of entropy associated with
the information on the holographic screen. Applying the first
principles, namely the holographic principle and the equipartition
law of energy, Verlinde derived the Newton's law of gravitation,
the Poisson equation and in the relativistic regime the Einstein
field equations. Although in \cite{Pad0} Padmanabhan observed that
the equipartition law of energy for the horizon degrees of freedom
combining with the thermodynamic relation $S=E/2T$, leads to the
Newton's law of gravity, however, the idea that gravity is not a
fundamental force and can be interpreted as the entopic force was
first pointed by Verlinde \cite{Ver}. Following \cite{Ver}, some
attempts have been done to investigate the entropic origin of
gravity in different setups (see
\cite{Cai4,Other,newref,sheyECFE,Ling,Modesto,Yi,Sheykhi2} and
references therein). Nevertheless, there are some critical
comments on Verlinde's proposal \cite{crit}. Strong criticism
against the entropic origin of gravity was presented by Visser
\cite{Vis} who claimed that the interpretation of gravity as an
entropic force is untenable. According to Visser arguments
\cite{Vis}, if one would like to reformulate classical Newtonian
gravity in terms of an entropic force, then the fact that
Newtonian gravity is described by a conservative force places
significant constraints on the form of the entropy and temperature
functions.

Although Verlinde's proposal has changed our understanding on the
origin and nature of gravity, but it considers the gravitational
field equations as the equations of emergent phenomenon and leave
the spacetime as a background geometric which has already exist.
Is it possible to regard the spacetime itself as an emergent
structure? Recently, by calculating the difference between the
surface degrees of freedom and the bulk degrees of freedom in a
region of space, Padmanabhan \cite{Pad1} argued that spacetime
dynamics can be emerged. As a result, he is able to explain the
origin of the acceleration of the universe expansion from his new
perspective \cite{Pad1}. According to Padmanabhan, the spatial
expansion of our universe can be regarded as the consequence of
emergence of space and \textit{the cosmic space is emergent as the
cosmic time progresses}. Using this new idea, Padmanabhan
\cite{Pad1} derived the Friedmann equation of a flat FRW Universe.
Following \cite{Pad1}, Cai obtained the Friedmann equation of a
higher dimensional FRW Universe in Einstein, Gauss- Bonnet and
Lovelock theory \cite{Cai1}. Similar derivation were also made by
the authors of \cite{Yang}. Instead of modifying the number of
degrees of freedom on the holographic surface of the Hubble
sphere, and the volume increase, the authors of \cite{Yang},
assumed that $(dV/dt)$ is
proportional to a function $f(\triangle N)$. Here $\triangle N=N_{\mathrm{sur%
}}-N_{\mathrm{bulk}}$, where $N_{\mathrm{sur}}$ is the number of
degrees of freedom on the boundary and $N_{\mathrm{bulk}}$ is the
number of degrees of freedom in the bulk. When the volume of the
spacetime is constant, the function $f(\triangle N)$ is equal to
zero. It is worth mentioning that the authors of \cite{Cai1,Yang}
only derived the Friedmann equations of the spatially flat FRW
Universe in Gauss-Bonnet and Lovelock gravities, and failed to
arrive at Friedmann equations with any spacial curvature in these
gravity theories. For this purpose, they proposed the Hawking
temperature associated with the Hubble horizon to be $T=H/2\pi$
and the volume of the universe is $V=4\pi H^{-3}/3$.\\ In this
paper, by modifying the original proposal of Padmanabhan
\cite{Pad1}, we are able to derive the Friedmann equation of the
FRW Universe with any spacial curvature. Note that in a nonflat
universe, the Hawking temperature and the volume are usually taken
as $T=1/2\pi \tilde{r}_A$ and $V=4\pi \tilde{r}^3_A/3$,
respectively, where $\tilde{r}_A$ is the apparent horizon radius
\cite{CaiKim}. We also generalize the study to the higher
dimensional spacetime and higher order gravities, and derive the
corresponding dynamical equations governing the evolution of the
universe with any spacial curvature not only in Einstein gravity,
but also in Gauss-Bonnet and more general lovelock gravity. For
consistency, in all cases we set the integration constant equal to
zero. In the next section we extract the Friedmann equation by
properly modifying the proposal of \cite{Pad1}. In section III, we
extend our study to higher order gravity theory in arbitrary
dimension. We summarize our results in section  IV.

\section{Friedmann equation in 4D Einstein gravity\label{Gen}}
We assume the background spacetime is spatially homogeneous and
isotropic which is described by the line element
\begin{equation}
ds^2={h}_{ab}dx^{a} dx^{b}+\tilde{r}^2(d\theta^2+\sin^2\theta
d\phi^2),
\end{equation}
where $\tilde{r}=a(t)r$, $x^0=t, x^1=r$, the two dimensional
metric is $h_{ab}$=diag $(-1, a^2/(1-kr^2))$. Here $k$ denotes the
curvature of space with $k = 0, 1, -1$ corresponding to flat,
closed, and open universes, respectively. The dynamical apparent
horizon, a marginally trapped surface with vanishing expansion, is
determined by the relation $h^{ab}\partial_{a}\tilde{r}
\partial_{b}\tilde{r}=0$. For a dynamical spacetime, the apparent horizon has
been argued to be a causal horizon and is associated with the
gravitational entropy and surface gravity \cite{Bak}. A simple
calculation gives the apparent horizon radius for the FRW Universe
as \cite{Hay}
\begin{equation}
\label{radius}
 \tilde{r}_A=\frac{1}{\sqrt{H^2+k/a^2}},
\end{equation}
where $H=\dot{a}/a$ is the Hubble parameter. It is widely accepted
that the apparent horizon is a suitable boundary of our universe,
from thermodynamic viewpoint, for which all laws of thermodynamics
are hold on it. Thermodynamical properties of the apparent horizon
has been studied in different setups \cite{SheyW1,Wang2,Shey3}.
 Following \cite{Pad1}, we assume the number of degrees of freedom
on the spherical surface of apparent horizon with radius
$\tilde{r}_A$ is proportional to its area and is given by
\begin{equation}
N_{\mathrm{sur}}=4S=\frac{4\pi \tilde{r}^2_A }{L_{p}^{2}},
\label{Nsur}
\end{equation}
where $L_{p}$ is the Planck length, $A=4\pi \tilde{r}^2_A$
represents the area of the apparent horizon and $S$ is the entropy
which obeys the area law. Assume the temperature associated with
the apparent horizon is the Hawking temperature \cite{CaiKim}
\begin{equation}\label{T}
T=\frac{1}{2\pi \tilde{r}_A},
 \end{equation}
and the energy contained inside the sphere with volume $V=4
\pi\tilde{r}^3_A/3$ is the Komar energy
\begin{equation}
E_{\mathrm{Komar}}=|(\rho +3p)|V.  \label{Komar}
\end{equation}
According to the equipartition law of energy, the bulk degrees of
freedom obey
\begin{equation}
N_{\mathrm{bulk}}=\frac{2|E_{\mathrm{Komar}}|}{T}.  \label{Nbulk}
\end{equation}
Through this paper we set $k_{B}=1=c=\hbar $ for simplicity. The
novel idea of Padmanabhan is that the cosmic expansion,
conceptually equivalent to the emergence of space, is being driven
towards holographic equipartition, and the basic law governing the
emergence of space must relate the emergence of space to the
difference between the number of degrees of freedom in the
holographic surface and the one in the emerged bulk \cite{Pad1}.
He proposed that in an infinitesimal interval $dt$ of cosmic time,
the increase $dV$ of the cosmic volume, in flat universe, is given
by
\begin{equation}
\frac{dV}{dt}=L_{p}^{2}(N_{\mathrm{sur}}-N_{\mathrm{bulk}}).  \label{dV}
\end{equation}
In general, one may expect ${dV}/{dt}$ to be some function of $(N_{\mathrm{%
sur}}-N_{\mathrm{bulk}})$ which vanishes when the latter does. In
this case one may regard Eq. (\ref{dV}) as a Taylor series
expansion of this function truncated at the first order
\cite{Pad1}. This approach was studied recently \cite{Yang}.

Motivated by (\ref{dV}), we propose the volume increase, in a
nonflat FRW Universe, is still proportional to the difference
between the number of degrees of freedom on the apparent horizon
and in the bulk, but the function of proportionality is not just a
constant, and it equals to the ratio of the apparent horizon and
Hubble radius. Therefore we write down
\begin{equation}
\frac{dV}{dt}=L_{p}^{2}\frac{\tilde{r}_A}{H^{-1}}
(N_{\mathrm{sur}}-N_{\mathrm{bulk}}). \label{dV1}
\end{equation}
It is well known that for pure de Sitter spacetime the number of
degrees of freedom in a bulk and the number of degrees of freedom
on the boundary surface are equal, namely $N_{\rm sur}=N_{\rm
bulk}$ \cite{Pad1}. Since our universe, is not exactly de Sitter
but it is asymptotically de Sitter, thus for our universe,
Padmanabhan proposed \cite{Pad1}
\begin{equation}
\frac{dV}{dt}\propto(N_{\mathrm{sur}}-N_{\mathrm{bulk}}).
\label{dV0}
\end{equation}
In order to arrive at the desired dynamical equations for the FRW
Universe, in Einstein gravity, he assumed the constant of
proportionality to be $L_{p}^2$. For a nonflat Universe and other
gravity theories, the assumption (\ref{dV}) does not work and we
found out that it should be modified as in Eq. (\ref{dV1}). One
may regard the assumption (\ref{dV1}) to the status of a postulate
and verify whether it can lead to the correct Friedmann equations
describing the evolution of the Universe. In this paper, we will
show that with this modification, we are able to extract the
Friedmann equations with any spacial curvature in Einstein,
Gauss-Bonnet and more general Lovelock gravity. This may justify
the correctness of our assumption in (\ref{dV1}). For spatially
flat universe, $\tilde{r}_A =H^{-1}$, and one recovers the
proposal (\ref{dV}).

Taking the time derivative of the cosmic volume $V=4
\pi\tilde{r}^3_A/3$, we have
\begin{equation}
\frac{dV}{dt}=4\pi \tilde{r}^2_A \dot{\tilde{r}}_A. \label{dVt}
\end{equation}
Substituting the cosmic volume $V$ and the temperature (\ref{T})
in Eq. (\ref{Nbulk}), we find the numbers of degrees of freedom in
the bulk as
\begin{equation}
N_{\mathrm{bulk}}=-\frac{16\pi^2
}{3}(\rho+3p)\tilde{r}_A^4.\label{Nbulk1}
\end{equation}
In order to have $N_{\rm bulk}>0$, we take $\rho+3p<0$
\cite{Pad1}. Substituting Eqs. (\ref{Nsur}), (\ref{dVt}),
(\ref{Nbulk1}) into (\ref{dV1}), we arrive at
\begin{equation}
4\pi \tilde{r}^2_A \dot{\tilde{r}}_A=L_{p}^{2}
\frac{\tilde{r}_A}{H^{-1}} \left[\frac{4\pi
 \tilde{r}^2_A}{L_{p}^{2}}+\frac{16\pi^2}{3}(\rho+3p)\tilde{r}^4_A\right]. \label{ddot1}
\end{equation}
Rearranging the terms we obtain
\begin{equation}
4\pi \tilde{r}^2_A \left(\dot{\tilde{r}}_A H^{-1}-
 \tilde{r}_A \right)=\frac{16\pi^2
 L_{p}^{2}}{3}(\rho+3p)\tilde{r}^5_A,
\label{ddot2}
\end{equation}
which can be simplified as
\begin{equation}
\tilde{r}^{-3}_A(\dot{\tilde{r}}_A H^{-1}-
 \tilde{r}_A)=\frac{4\pi L_{p}^{2}}{3}\left[3(\rho+p)-2\rho\right].
\label{ddot3}
\end{equation}
Using the continuity equation, $\dot{\rho}+3H(\rho +p)=0,$ we
reach
\begin{equation}
\tilde{r}^{-3}_A(\dot{\tilde{r}}_A H^{-1}-
 \tilde{r}_A)=-\frac{4\pi L_{p}^{2}}{3}\left[\dot{\rho}H^{-1}+2\rho\right].
\label{ddot4}
\end{equation}
Multiplying the both hand side of (\ref{ddot4}) by factor
$2\dot{a}a$, and using the fact that $H^{-1}=a/\dot{a}$, we get
\begin{equation}
2\dot{a}a\tilde{r}^{-2}_A-2a^2 \dot{\tilde{r}}_A
\tilde{r}^{-3}_A=\frac{8\pi L_{p}^{2}}{3}\left[\dot{\rho}a^2+2\rho
\dot{a}a\right]. \label{ddot5}
\end{equation}
The above equation can be further rewritten as
\begin{equation}
\frac{d}{dt}\left(a^2
\tilde{r}_A^{-2}\right)=\frac{d}{dt}\left[a^2\left(H^2+\frac{k}{a^2}\right)\right]=\frac{8\pi
L_{p}^{2}}{3}\frac{d}{dt}(\rho a^2), \label{Fr1}
\end{equation}
where we have also used relation (\ref{radius}). Integrating, we
obtain
\begin{equation}
H^{2}+\frac{k}{a^{2}}=\frac{8\pi L_{p}^{2}}{3}\rho ,  \label{Fr2}
\end{equation}
where we have set the integration constant equal to zero. In this
way we derive the Friedmann equation of the FRW Universe with any
spacial curvature, by calculating the difference between the
number of degrees of freedom in the bulk and on the apparent
horizon. Let us stress here the difference between our derivation
and ones presented in \cite{Cai1,Yang}. The authors of
\cite{Cai1,Yang} arrived at (\ref{Fr2}), by using proposal
\cite{Pad1} given by Eq. (\ref{dV}), and interpreting the
integration constant as the special curvature, while we arrive at
the same result by modifying the proposal of \cite{Pad1}  in the
form of (\ref{dV1}), and setting the integration constant equal to
zero.
\section{Friedmann equation in Gauss-Bonnet and Lovelock gravity\label{entr}}
In this section, we apply the approach developed in the previous
section to derive the Friedmann equations in Gauss-Bonnet and more
general Lovelock gravity with any spacial curvature. This is the
first derivation of Friedmann equations in these gravity theories
in a nonflat FRW Universe by using the novel idea presented in
\cite{Pad1}. We first extend the approach of the previous section
to the $(n+ 1)$-dimensional spacetime. In this case the number of
degrees of freedom on the apparent horizon turn out to be
\cite{Cai1}
\begin{equation}
N_{\mathrm{sur}}=\alpha \frac{A}{L_{p}^{2}}, \label{Nsur3}
\end{equation}
where $A=n \Omega_n \tilde{r}^{n-1}_A $ and $\alpha=(n-1)/2(n-2)$,
with $\Omega_n$ is the volume of an unit $n$-sphere. We also
modify  our proposal in (\ref{dV1}) a little as
\begin{equation}
\alpha \frac{dV}{dt}=L_{p}^{n-1}\frac{\tilde{r}_A}{H^{-1}}
(N_{\mathrm{sur}}-N_{\mathrm{bulk}}), \label{dV2}
\end{equation}
where the volume of the $n$-sphere is $V =\Omega_n
\tilde{r}^{n}_A$. The bulk Komar energy in $(n+1)$-dimensions is
given by \cite{Cai2}
\begin{equation}
E_{\rm Komar}=\frac{(n-2)\rho+np}{n-2}V, \label{Ek2}
\end{equation}
and hence the bulk degrees of freedom is obtained as
\begin{equation}
N_{\rm bulk}=-4 \pi \Omega_n \tilde{r}^{n+1}_A
\frac{(n-2)\rho+np}{n-2}, \label{Nbulk3}
\end{equation}
where we take $(n-2)\rho+np<0$ in order to have $N_{\rm bulk}>0$
\cite{Pad1}. Substituting Eqs. (\ref{Nsur3}) and (\ref{Nbulk3}) in
relation (\ref{dV2}), one gets
\begin{equation}
\tilde{r}^{-2}_A-\dot{\tilde{r}}_A H^{-1} \tilde{r}^{-3}_A
=-\frac{8 \pi L_{p}^{n-1}}{n(n-1)}[(n-2)\rho+np]. \label{Fr3}
\end{equation}
Multiplying the both hand side by factor $2\dot{a}a$, after using
the continuity equation in $(n+1)$-dimensions as
\begin{equation}
\dot{\rho}+nH(\rho +p)=0,  \label{cont2}
\end{equation}
we arrive at
\begin{equation}
\frac{d}{dt}\left[a^2
\left(H^2+\frac{k}{a^2}\right)\right]=\frac{16 \pi
L_{p}^{n-1}}{n(n-1)} \frac{d}{dt}(\rho a^2). \label{Fr4}
\end{equation}
Integrating, we find
\begin{equation}
H^2+\frac{k}{a^2}=\frac{16 \pi L_{p}^{n-1}}{n(n-1)}\rho,
\label{Fr5}
\end{equation}
where we have set the integration constant equal to zero. This is
the Friedmann equation of $(n+1)$-dimensional FRW Universe with
any spacial curvature \cite{CaiKim}.

Up to now we only considered Einstein gravity, and derive the
corresponding Friedmann equations in a Universe with spacial
curvature. Now we want to see whether the above procedure works or
not in other gravity theories such as the Gauss-Bonnet and more
general Lovelock gravity. Lovelock gravity is the most general
lagrangian which keeps the field equations of motion for the
metric of second order, as the pure Einstein-Hilbert action
\cite{Lov}. Let us first consider the Gauss-Bonnet theory. The key
point which should be noticed here is that in Gauss-Bonnet gravity
the entropy of the holographic screen does not obey the area law.
Static black hole solutions of Gauss-Bonnet gravity have been
found and their thermodynamics have been investigated in ample
details \cite{Bou,caigb}. The entropy of the static spherically
symmetric black hole in Gauss-Bonnet theory has the following
expression \cite{caigb}
\begin{equation}\label{Sbh}
S=\frac{A_{+}}{4L_{p}^{n-1}}\left[1+\frac{n-1}{n-3}\frac{2
\tilde{\alpha}}{r_{+}^2}\right],
\end{equation}
where $A_{+}=n \Omega_n r^{n-1}_{+}$ is the horizon area and $r_+$
is the horizon radius. In the above expression
$\tilde{\alpha}=(n-2)(n-3)\alpha$, where $\alpha$ is the
Gauss-Bonnet coefficient which is positive \cite{Bou}. For $n=3$
we have $\tilde{\alpha}=0$, thus the Gauss-Bonnet correction term
contributes only for $n\geq 4$. We assume the entropy expression
(\ref{Sbh}) also holds for the apparent horizon of the FRW
Universe in Gauss-Bonnet gravity. The only change we need to apply
is the replacement of the horizon radius $r_+$ with the apparent
horizon radius $\tilde{r}_A$, namely
\begin{equation}\label{S2}
S=\frac{A}{4L_{p}^{n-1}}\left[1+\frac{n-1}{n-3}\frac{2
\tilde{\alpha}}{\tilde{r}_A^2}\right],
\end{equation}
where $A=n \Omega_n \tilde{r}_A^{n-1}$ is the apparent horizon
area. We define the effective area of the holographic surface
corresponding to the entropy (\ref{S2}) as
\begin{eqnarray}
\widetilde{A} = n \Omega_n
\tilde{r}_A^{n-1}\left[1+\frac{n-1}{n-3}\frac{2
\tilde{\alpha}}{\tilde{r}_A^2}\right].
\end{eqnarray}
Now we calculate the increasing in the effective volume as
\begin{eqnarray}\label{dVt1}
\frac{d\widetilde{V}}{dt}&=&\frac{\tilde{r}_A}{(n-1)}\frac{d\widetilde{A}}{dt}=n
\Omega_n \dot{\tilde{r}}_A \tilde{r}^{n-1}_A(1+2 \tilde
{\alpha}\tilde{r}^{-2}_A)\\ &=& -\frac{n\Omega_n
\tilde{r}^{n+2}_A}{2}\frac{d}{dt}\left(\tilde{r}^{-2}_A+\tilde{\alpha}
\tilde{r}^{-4}_A\right). \label{dVt2}
\end{eqnarray}
Inspired by (\ref{dVt2}), we propose that the number of degrees of
freedom on the apparent horizon, in Gauss-Bonnet gravity, is given
by
\begin{equation}
N_{\mathrm{sur}}=\frac{\alpha n\Omega_n \tilde{r}^{n+1}_A}{
L_{p}^{n-1}}\left(\tilde{r}^{-2}_A+\tilde{\alpha}
\tilde{r}^{-4}_A\right).  \label{Nsur2}
\end{equation}
The bulk degrees of freedom is still given by (\ref{Nbulk3}).
Inserting Eqs. (\ref{Nbulk3}), (\ref{dVt1}) and (\ref{Nsur2}) in
relation (\ref{dV2}), with replacing $V\rightarrow \tilde{V}$, we
obtain
\begin{eqnarray}
&&(\tilde{r}^{-2}_A+\tilde{\alpha}\tilde{r}^{-4}_A)-\dot{\tilde{r}}_A
H^{-1} \tilde{r}^{-3}_A(1+2\tilde{\alpha}\tilde{r}^{-2}_A)\\
&=&-\frac{8 \pi L_{p}^{n-1}}{n(n-1)}[(n-2)\rho+np]. \label{Frgb1}
\end{eqnarray}
Multiplying the both hand side of (\ref{Frgb1}) by factor
$2\dot{a}a$, with help of continuity equation (\ref{cont2}) and
relation (\ref{radius}), we get
\begin{equation}
\frac{d}{dt}\Bigg{\{}a^2 \left[H^2+\frac{k}{a^2}+\tilde{\alpha}
\left(H^2+\frac{k}{a^2}\right)^2\right]\Bigg{\}}=\frac{16 \pi
L_{p}^{n-1}}{n(n-1)} \frac{d}{dt}(\rho a^2). \label{Frgb2}
\end{equation}
Integrating, we find
\begin{equation}
H^2+\frac{k}{a^2}+\tilde{\alpha}
\left(H^2+\frac{k}{a^2}\right)^2=\frac{16 \pi L_{p}^{n-1}}{n(n-1)}
\rho, \label{Frgb3}
\end{equation}
where again we have set the integration constant equal to zero.
This is indeed, the corresponding Friedmann equation of the FRW
Universe with any spacial curvature in Gauss-Bonnet gravity
\cite{CaiKim}. Note that the authors of Refs. \cite{Cai1,Yang}
could derive the above equation only in a flat FRW Universe, while
we derive it with arbitrary spacial curvature. This may show the
viability of our proposal (\ref{dV2}).

Finally, we consider the more general Lovelock gravity. The
entropy of the spherically symmetric black hole solutions in
Lovelock theory can be expressed as \cite{caiLo}
\begin{equation}\label{SL}
S=\frac{A_{+}}{4 L_{p}^{n-1}}\sum_{i=1}^m
\frac{i(n-1)}{(n-2i+1)}{\hat{c}_{i}}{{r}_{+}}^{2-2i},
\end{equation}
where $m=[n/2]$ and the coefficients ${\hat{c}_{i}}$ are given by
\begin{equation}\label{constant}
{\hat{c}_{0}}=\frac{{c_{0}}}{n(n-1)}, \ \  {\hat{c}_{1}}=1, \ \
{\hat{c}_{i}}=c_{i}\prod_{j=3}^{2m}(n+1-j) \  \  i>1.
\end{equation}
We further assume the entropy expression (\ref{SL}) are valid for
a FRW Universe bounded by the apparent horizon in the Lovelock
gravity provided we replace the horizon radius $r_{+}$ with the
apparent horizon radius $\tilde{r}_A$, namely
\begin{equation}\label{mSL}
S=\frac{A}{4 L_{p}^{n-1}}\sum_{i=1}^m
\frac{i(n-1)}{(n-2i+1)}{\hat{c}_{i}}{\tilde{r}_A}^{2-2i}.
\end{equation}
It is easy to show that, the first term in the above expression
leads to the well known area law. The second term yields the
apparent horizon entropy in Gauss-Bonnet gravity. We suppose from
the entropy expression that the effective area of the apparent
horizon in Lovelock gravity is given by
\begin{eqnarray}
\widetilde{A} = n\Omega_n \tilde{r}^{n-1}_A \sum_{i=1}^m
\frac{i(n-1)}{(n-2i+1)}{\hat{c}_{i}}{\tilde{r}_A}^{2-2i},
\end{eqnarray}
and the increase of the effective volume is then given by
\begin{eqnarray}\label{dVtL1}
\frac{d\widetilde{V}}{dt}&=&\frac{\tilde{r}_A}{(n-1)}\frac{d\widetilde{A}}{dt}
=n \Omega_n  \tilde{r}^{n+1}_A\left(\sum_{i=1}^m
 i\hat{c}_{i}{\tilde{r}_A}^{-2i}\right) \dot{\tilde{r}}_A\\ &=& -\frac{n\Omega_n
\tilde{r}^{n+2}_A}{2}\frac{d}{dt}\left(\sum_{i=1}^m
 \hat{c}_{i}{\tilde{r}_A}^{-2i}\right). \label{dVtL2}
\end{eqnarray}
In this case,  we assume from (\ref{dVtL2}) that the number of
degrees of freedom on the apparent horizon, in Lovelock gravity,
is
\begin{equation}
N_{\mathrm{sur}}=\frac{\alpha n\Omega_n }{
L_{p}^{n-1}}\tilde{r}^{n+1}_A\sum_{i=1}^m
 \hat{c}_{i}{\tilde{r}_A}^{-2i}.  \label{NsurL}
\end{equation}
Substituting (\ref{Nbulk3}), (\ref{dVtL1}) and (\ref{NsurL}) into
(\ref{dV2}), we reach
\begin{eqnarray}
&&\sum_{i=1}^m
 \hat{c}_{i}{\tilde{r}_A}^{-2i}-\dot{\tilde{r}}_A
H^{-1} \sum_{i=1}^m
 i\hat{c}_{i}{\tilde{r}_A}^{-2i-1}\\ &=&-\frac{8 \pi L_{p}^{n-1}}{n(n-1)}[(n-2)\rho+np].
\label{FrL1}
\end{eqnarray}
Multiplying the both hand side by factor $2\dot{a}a$, after using
the continuity equation (\ref{cont2}) as well as definition
(\ref{radius}), we obtain
\begin{equation}
\frac{d}{dt} \left[a^2\sum_{i=1}^m
 \hat{c}_{i}\left(H^2+\frac{k}{a^2}\right)^{i}\right] =\frac{16 \pi
L_{p}^{n-1}}{n(n-1)} \frac{d}{dt}(\rho a^2). \label{FrL2}
\end{equation}
After integrating and setting the  constant of integration equal
to zero, we find the corresponding Friedmann equation of the FRW
Universe with any spacial curvature in Lovelock gravity,
\begin{equation}
\sum_{i=1}^m
 \hat{c}_{i}\left(H^2+\frac{k}{a^2}\right)^{i}=\frac{16 \pi
L_{p}^{n-1}}{n(n-1)}\rho. \label{FrL3}
\end{equation}
This is exactly the result obtained in \cite{CaiKim} by applying
the first law of thermodynamics on the apparent horizon of the FRW
Universe in Lovelock gravity. Here we arrived at the same result
by using quite different approach. This indicates that, given the
entropy expression at hand, one is able to reproduce the
corresponding dynamical equation with any spacial curvature, by
applying the proposal (\ref{dV2}).
\section{Summary and discussion\label{con}}
We have investigated the novel idea recently proposed by
Padmanabhan \cite{Pad1}, which states that the emergence of space
and Universe expansion can be understood by calculating the
difference between the number of degrees of freedom on the Hubble
horizon and the one in the emerged bulk. Applying this idea to a
flat FRW Universe with Hubble horizon, he derived the dynamical
equation describing the evolution of the Universe \cite{Pad1}. In
this paper, by properly modification his idea, we derived the
Friedmann equation of a FRW Universe with any spacial curvature.
Our approach not only works in Einstein gravity, but also works
very well in Gauss-Bonnet and more general Lovelock gravity. The
key assumption here is that in a nonflat Universe, the volume
increase, is still proportional to the difference between the
number of degrees of freedom on the apparent horizon and in the
bulk, but the function of proportionality is not just the constant
$L_p^2$, instead it equals to the ratio of the apparent horizon
radius and the Hubble radius, i.e., $L_p^2\tilde{r}_A/H^{-1}$.

It is important to note that Padmanabhan's proposal (\ref{dV}) can
lead to the Friedmann equation with spacial curvature only in
Einstein gravity \cite{Cai1,Yang}. The main result of the present
work is that the modified proposal (\ref{dV1}) can lead to the
Friedmann equations of the FRW Universe with any spacial curvature
in higher order gravity theories. Indeed, while the authors of
\cite{Cai1,Yang} interpreted the integration constant as the
spatial curvature $k$ in Einstein gravity, they failed to
interpret the constant of integration as the spatial curvature in
the cases of Gauss-Bonnet and Lovelock gravities. This is due to
the fact that, in Einstein gravity, the de Sitter Universe can be
described either by $k=0$ or $k=1$. As a result, in Gauss-Bonnet
and Lovelock gravity, with proposal (\ref{dV}), they could only
derive the Friedmann equations of the flat Universe.

In summary, given the entropy expression at hand, one is able to
reproduce the corresponding dynamical equation of the FRW Universe
with any spacial curvature, by calculating the difference between
the horizon degrees of freedom and the bulk degrees of freedom in
a region of space and applying the proposal (\ref{dV1}). The
results obtained in this paper together with those of
\cite{Cai1,Yang} further support the new proposal of Padmanabhan
\cite{Pad1} and its modification as (\ref{dV1}) and show that this
approach is powerful enough to apply for deriving the dynamical
equations describing the evolution of the Universe in other
gravity theories with any spacial curvature.

\acknowledgments{I thank the referee for constructive comments
which helped me to improve the paper significantly. I also thank
the Research Council of Shiraz University. This work has been
supported financially by Center for Excellence in Astronomy and
Astrophysics of IRAN (CEAAI-RIAAM).}


\begin{thebibliography}{99}
\bibitem{Jac}  T. Jacobson, Phys. Rev. Lett. {\bf75}, 1260 (1995).
\bibitem{Ver}  E. Verlinde, JHEP {\bf1104}, 029 (2011).

\bibitem{Pad0}  T. Padmanabhan, Mod. Phys. Lett. A {\bf25}, (2010) 1129.

\bibitem{crit} S. Gao, arXiv : 1002.2668; H. Culetu, arXiv:1002.3876; A. Kobakhidze, arXiv:1009.5414; M. Chaichian, M. Oksanen and A. Tureanu,
arXiv:1104.4650.

\bibitem{Vis}  M. Visser, JHEP {\bf1110}, 140 (2011).


\bibitem{Cai4} R.G. Cai, L. M. Cao and N. Ohta, Phys. Rev. D {\bf81}, 061501 (2010).

\bibitem{Other}  R.G. Cai, L. M. Cao and N. Ohta, Phys. Rev. D {\bf81}, 084012 (2010);
\\ Y.S. Myung, Y.W Kim, Phys. Rev. D {\bf81}, 105012 (2010);
\\ R. Banerjee, B. R. Majhi, Phys. Rev. D {\bf81}, 124006 (2010);
\\ S.W. Wei, Y. X. Liu, Y. Q.  Wang, Commun. Theor. Phys. {\bf56}, 455 (2011);
\\ Y. X. Liu, Y. Q.  Wang, S. W. Wei, Class. Quantum Gravit. {\bf27}, 185002 (2010);
\\ R.A. Konoplya, Eur. Phys. J. C {\bf69}, 555 (2010);
\\ H. Wei, Phys. Lett. B {\bf692}, 167 (2010).


\bibitem{newref}  C. M. Ho, D. Minic and Y. J. Ng, Phys.\ Lett.\ B {\bf 693}, 567
(2010);\\
V.V. Kiselev, S.A. Timofeev  Mod. Phys. Lett. A {\bf26}, (2011) 109;\\
W. Gu, M. Li and R. X. Miao,  Sci.China G {\bf54}, 1915 (2011),
arXiv:1011.3419;\\ R. X. Miao, J. Meng and M. Li, Sci. China G
{\bf55}, 375 (2012), arXiv:1102.1166.


\bibitem{sheyECFE} A. Sheykhi,  Phys. Rev. D {\bf81}, 104011 (2010).

\bibitem{Ling} Y. Ling and J.P. Wu, JCAP {\bf1008}, (2010), 017.

\bibitem{Modesto} L. Modesto,  A. Randono, arXiv:1003.1998;
\\ L. Smolin, arXiv:1001.3668;\\ X. Li,  Z. Chang,
arXiv:1005.1169.

\bibitem{Yi} Y.F. Cai, J. Liu, H. Li, Phys. Lett. B {\bf690}, (2010) 213;
\\ M. Li and Y. Wang, Phys. Lett. B {\bf687}, 243 (2010).

\bibitem{Sheykhi2} S. H. Hendi and A. Sheykhi, Phys.  Rev.  D {\bf83}, 084012 (2011);
\\ A. Sheykhi and S. H. Hendi, Phys.  Rev.  D {\bf84}, 044023 (2011);
\\ S. H. Hendi and A. Sheykhi, Int. J. Theor. Phys. {\bf51}, 1125 (2012)
;\\ A. Sheykhi and Z. Teimoori, Gen Relativ Gravit. {\bf44}, 1129
(2012);\\ A. Sheykhi,  Int. J. Theor. Phys. {\bf51}, 185 (2012);\\
A. Sheykhi, K. Rezazadeh Sarab, JCAP {\bf10}, 012 (2012).




\bibitem{Pad1}  T. Padmanabhan, arXiv:1206.4916.

\bibitem{Cai1}  R. G. Cai, JHEP {\bf11}, 016  (2012).
\bibitem{Yang}  K. Yang, Y. X. Liu and Y. Q. Wang, Phys. Rev. D {\bf86}, 104013
(2012).

\bibitem{CaiKim} R. G. Cai and S. P. Kim, JHEP {\bf0502}, 050
(2005).


\bibitem{Bak} D. Bak and S. J. Rey, Class. Quantum Gravit. {\bf17}, L83 (2000);
\\ S. A. Hayward, S. Mukohyama and M. C. Ashworth, Phys. Lett. A
{\bf256}, 347 (1999).

\bibitem{Hay} S. A. Hayward, Class. Quantum Gravit. {\bf15}, 3147 (1998).

\bibitem{SheyW1}  A. Sheykhi, B. Wang and R. G. Cai, Nucl. Phys. B \textbf{
779}, 1 (2007);\\ A. Sheykhi, B. Wang and R. G. Cai, Phys. Rev. D
\textbf{76}, 023515 (2007);\newline A. Sheykhi, JCAP \textbf{05},
019 (2009);\\A. Sheykhi, Class. Quantum Gravit. \textbf{27},
025007 (2010);\\ A. Sheykhi, Eur. Phys. J. C \textbf{69}, 265
(2010).

\bibitem{Wang2} J. Zhou, B. Wang, Y. Gong, E. Abdalla,  Phys. Lett. B {\bf652}, 86 (2007).

  \bibitem{Shey3} A. Sheykhi, B. Wang, Phys. Lett. B {\bf678}, (2009) 434\\ A. Sheykhi, B.
Wang, Mod. Phys. Lett. A,  Vol. {\bf25}, No. 14, 1199 (2010).

\bibitem{Cai2} R. G. Cai, L. M. Cao and N. Ohta, Phys. Rev. D {\bf81}, 061501
(2010).

\bibitem{Lov} D. Lovelock, J. Math. Phys. (N.Y.) {\bf12}, 498 (1971).

\bibitem{Bou} D. G. Boulware and S. Deser, Phys. Rev. Lett. {\bf55}, 2656
(1985);\\
J. T. Wheeler, Nucl. Phys. B {\bf268}, 737 (1986);\\ Nucl. Phys. B
{\bf273}, 732 (1986);\\ R. C. Myers and J. Z. Simon, Phys. Rev. D
{\bf38}, 2434 (1988).
\bibitem{caigb} R. G. Cai, Phys. Rev. D {\bf65}, 084014 (2002);\\ R. G. Cai and Q. Guo,
Phys. Rev. D {\bf69}, 104025 (2004);\\ R. G. Cai and K. S. Soh,
Phys. Rev. D {\bf59}, 044013 (1999).

\bibitem{caiLo}  R. G. Cai, Phys. Lett. B {\bf582}, 237 (2004).
\end{thebibliography}
\end{document}